# On the Possibility of Reversible Magnonic Logic Gates

Alexander Khitun

Electrical Engineering Department, University of California at Riverside, Riverside, California, 92521

**Abstract**

We propose and develop a concept of magnonic logic gates enabling reversible computing. The gates consist of passive elements: waveguides, cross-junctions and phase shifters. Logical 0 and 1 are encoded in the relative phase of the propagating spin wave packets (0 or $\pi$). The gates contain several possible trajectories for each packet to propagate from the input to the output. Re-direction of the spin wave packets among the possible trajectories is due to the interference in the magnetic cross-junctions. Two wave packets coming to the cross-junction in-phase propagate through the junction without reflection. Two packets coming out-of-phase to the junction are completely reflected back. The operation of the cross-junction is illustrated by numerical modeling. We estimate the power dissipation in the proposed circuits and the feasibility of cascading such magnetic devices in large circuits. The proposed gates may potentially provide a route to magnetic reversible logic circuitry with power dissipation less than *kT* per operation.

**I. Introduction**

Power dissipation has emerged as one of the main obstacles to further increase of the computational throughput [1]. In the past four decades, a straightforward approach to computation performance improvement was associated with the increase of the number of transistors and speeding up the switching of the individual transistor. According to the Moore's law [2], the number of transistors in an integrated circuit doubled every two years. Meanwhile the switching frequency increased to the GHz range. Both these trends lead to increased dissipated power. Today the power density dissipated in the chip active area is of the order of 100W/cm$^2$, which approaches the limit for the conventional air-cooling. The use of additional cooling mechanisms (e.g. water cooling, thermoelectric cooling) is associated with significant technological complications and may provide only a temporary solution. This fact stimulates a great deal of interest to novel computational paradigms able to overcome the power dissipation problem.

There are two fundamental reasons for power dissipation in the conventional computational devices: logic irreversibility of the exploited logic gates and physical irreversibility of the basic elements – transistors. Most of basic logic gates are logically irreversible. For example, AND and OR are the three-terminal gates with two inputs and one output. The process of computation in these gates is logically irreversible, as it is not possible to reconstruct two inputs from a single output. Thus, every step of computation is accompanied by a loss of one bit of information. As it was pointed out by Landauer [3], each lost bit is equivalent to at least $kT\ln 2$ energy loss, where $k$ is the Boltzmann's constant and $T$ is the operation temperature. It would be naive to believe that simply using two-input two-output logic gates may solve the problem of power dissipation. The logic gates consist of transistors and each of the transistors dissipates power generating Joule's heat. For the current technology, the energy dissipated by a transistor per switching event is of the order of $10^5 kT$ [1], which is several orders of magnitude above the Landauer's limit. Though, there is still a room for energy loss minimization in the conventional transistor-based circuits, the radical solution to the power dissipation problem is in the utilization of logically reversible gates consisting of physically reversible elements.

Quantum computing is an example of physically reversible computing, where the process of computation is associated with the evolution of the quantum-mechanical system. Theoretically, there is no power dissipation during the computational process until the final result is read out. Besides, proponents of quantum computers claim tremendous improvements over traditional computing methods for certain types of problems, such as the prime factorization[4]. The advantages of quantum computer originate from the utilization of quantum state superposition and coherent manipulation of entangled states. There are a few working prototypes with several devices designed to function as quantum bits (qubits), which have been practically demonstrated[5-7]. However, quantum computer is still far from practical application because requirements of preservation of entanglement in these qubits are still not fulfilled.

An important question to ask is whether it possible to realize reversible and low-power logic circuits by utilizing classical waves. It proves to be possible to implement certain algorithms associated with quantum computing using classical wave interference techniques. It has been reported that classical wave interference can implement algorithms where quantum entanglement is not required (e.g. Deutsch-Jozsa algorithm [8], Grover Search algorithm [9], Bernstein-Vazirani algorithm [10]). The latter opens an intriguing possibility to build a new class of classical wave-based logic devices with capabilities intermediate between the conventional transistor-based and purely quantum computers.

In this work, we analyze the possibility of making wave-based magnetic reversible logic gates, where the process of computation is associated with the phase change of a propagating wave. In Section II, we describe the principle of operation of reversible magnonic logic devices. In Section III, we present the results of numerical simulations illustrating the operation of a magnonic cross-junction. In Section IV, we analyze the power dissipation in the magnonic logic gate operating at a finite temperature and discuss the problems and shortcomings of the proposed devices.

**II. Principle of operation**

A reversible wave-based logic device (shown in Fig.1(a)) consists of passive elements: waveguides, junctions, and phase shifters. Inputs and outputs are classical wave packets. All packets are assumed to have identical frequency spectra and the same amplitudes. A bit of information is encoded as a relative phase with respect to the reference packet as shown in Fig.1. It should be clarified that each of the wave packets may envelope a large number of waves of different frequencies with a very short temporal coherency. However, the phase difference between the waves of the same frequency in distinct packets may be preserved for much longer time (longer than the time required for computation). Two relative phases 0 and $\pi$ correspond to logic states 0 and 1, respectively. Passing through the phase shifter, the wave packet obtains an additional $\pi$-phase shift, which corresponds to a logic inversion (e.g. 0 to 1, 1 to 0). Two simplest examples of phase-based logic gates are illustrated in Fig.1(b). The buffer gate is a waveguide without a phase shifter. The relative phase between the propagating packet and the reference packet remains the same. Logic Inverter is a waveguide with a $\pi$-phase shifter.

There are several possible ways for each wave packet to travel from the input to the output in the gate shown in Fig.1(A). Some of the possible trajectories contain phase shifters and others do not. The particular way of packet propagation depends on the propagation through the waveguide cross-junctions where two packets interfere. Depending on the relative phase of the incoming wave packets, the packets may come through the junction (constructive interference) or may be reflected from the junction (destructive interference). Later on in this work, we present the results of numerical modeling illustrating spin wave transport through a cross-like junction. An elementary act of computation in the described scheme is associated with the redirection of a wave packet. The result of computation is determined by measuring the relative phases between the output packets and the reference packet. In some aspects, the described logic functionality is reminiscent of the billiard-ball computer proposed by Fredkin and Toffoli [11], where billiard balls move through a specially designed structure. In that scheme, a ball may be scattered due to a collision with the structure and a collision with other balls. As a result of a number of collisions, the balls change their trajectories. The latter is treated as computation.

In general, the described approach can be applied to the different types of waves. We consider spin waves packets as one possible approach convenient for practical realization. Spin wave is a collective oscillation of spins in the magnetic lattice, which can propagate only in magnetic materials. This makes it easy to build magnetic waveguides and to guide spin wave propagation. The redirection of spin wave packets can be done by a magnetic cross junction. In contrast to electro-magnetic waves, a propagating spin wave has two components for magnetization: along and perpendicular to the direction of propagation as illustrated in Fig.2. The waves propagating in two perpendicular directions may have the same or opposite magnetization components leading to constructive or destructive interference in the junction. The packets will propagate through the junction in case of constructive interference, and will be scattered back in case of destructive interference. There is a variety of solutions for a magnonic phase shifter. Spin wave dispersion depends on the waveguide material, geometry, and an external magnetic field. An additional phase shift can be obtained at any desired part of the magnetic waveguide by changing the waveguide thickness, width, or by using pinned layers providing a magnetic field. In the most compact case, a π-phase shift can be provided by a special-shape domain wall as described in [12].

### III. Numerical modeling

Magnetic cross-junction is the key element in the proposed scheme. In order to illustrate its operation, we present the results of numerical simulations. We consider the two perpendicular chains of spins as shown in Fig. 2. The neighboring spins in the magnetic wires are coupled via exchange interaction, so the Hamiltonian of the system has the following form [13]:

$$H = -J \sum_{j\delta} S_j S_{j+\delta} - 2\mu H_0 \sum_j S_{jz}, \qquad (1)$$

where $J$ is the exchange coupling constant with the dimension of energy, $S_j$ and $S_{j+\delta}$ are the electron-spin operators, $S_{jz}$ is the spin projection along the z direction, the index $\delta$ runs over nearest neighbors of spin $j$, $\mu$ is the magnetic moment, $H_0$ is the external magnetic field strength. The evolution equation for spin $j$ takes the following forms:

$$\hbar \frac{d\vec{S}_j}{dt} = \vec{\mu} \times \vec{B}_j, \qquad (2)$$

where $B_j$ is the effective magnetic field induction acting on spin $j$, which arises from the sum of the exchange field due to the coupling with the nearest neighbor spins and the external magnetic field. The detailed explanations on the one-dimensional chain model can be found elsewhere [13]. In our case, two chains intersect in just one point - one spin (depicted by subscript $p$ in Fig.2). The purpose of our simulations is to illustrate the transmission/ reflection of two spin wave packets propagating through the junction.

At the initial moment of time, we assume all spins to be directed along the Z axis and parallel to the external magnetic field $S_z=1$, $S_x=S_y=0$. Then, we introduce a small perturbation to the edge spins in the chains (points A and B in Fig.2). The perturbation is a small in-plane deviation $S_x$ (perpendicular to the external field) of the spin $S_x/S$~0.05. The perturbations result in collective oscillations of spins in each magnetic wire - spin wave packets. The packets propagate through the magnetic wires towards the intersection point. In Fig. 3, we present the results of numerical simulations showing packet propagation through the junction. We considered two cases: (i) $Sx_A=Sx_B$, and (ii) $Sx_A=-Sx_B$. There are several snapshots showing the normalized projection $S_x/S$ for the spins near the junction. In the first case, the packets propagate through the junction with no reflection. In the second case, two packets are completely reflected from the junction. The physical explanation is pretty simple. By introducing the same initial deviation to the edge spins (first case, $Sx_A=Sx_B$), we excite two spin wave packets with the same initial phase. The packets interfere constructively at the point of intersection and propagate through the junction. In the second case, the packets have a π-phase shift. Approaching the intersection point, the effective magnetic fields produced by the nearest-neighbor spins in two chains tend to rotate the spin at the intersection in the opposite directions. As a result of this competition (destructive interference), the spin in the intersection stays still and both packets are reflected back.

In Fig. 3, we show an example of the two-bit reversible logic gate comprising one cross-junction and one phase shifter (e.g. a domain wall). There are two input/output ports marked by letters A and B. The principle of operation is as follows. Two input spin wave signals are excited at points A and B as shown in Fig.3. The input signals are spin wave packets, which may have two initial relative phases 0 or $\pi$ with respect to the control signal, corresponding to logic states 0 and 1, respectively. The input spin waves propagate from the points of excitation toward the cross-junction. Then, there are two possible scenarios for spin wave transport through the junction depending on the relative phases. (i) The waves completely reflected from the junction if they have a $\pi$ relative phase (e.g. one wave has a 0 phase difference with the reference signal, and the other one has a $\pi$ phase shift with respect to the reference signal). In this case, the waves return to the excitation points without any change of the relative phase. The latter corresponds to the logic operation 01→01, and 10→10. (ii) The waves propagate through the junction without reflection if they have the same initial phase (both 0 or both $\pi$ with respect to the reference signal). In this case, the wave propagating towards the A point passes through the phase shifter and gains a $\pi$-phase shift. The wave propagating towards the B point preserves the relative phase with respect to the reference signal. This situation corresponds to the logic transition 00→10, and 11→01. The truth table is displayed in Fig. 3. The presented example is aimed to show the possibility of building physically and logically reversible logic gates by exploiting wave interference and phase shifters.

## IV. Discussion

The principle of operation of the described magnonic logic circuits is fundamentally different from the conventional transistor circuit operation. Logic state is encoded into the phase of the propagating wave packet and computation is accomplished via the packet re-direction. The gates consist of only passive components, which do not require any external energy source for operation. The propagation of the bit-carrying packets is time-reversible. From the theoretical point of view, these circuits may dissipate zero energy per operation. That would be possible for the ideal cross-junctions enabling 100% transmission/reflection and for the ideal defect-free waveguides operating at zero temperature. In reality, there will be inevitable power dissipation due to the following reasons: (i) non-reversible losses due to the scattering processes (e.g. magnon-magnon and magnon-phonon scattering, scattering on structure imperfections), and (ii) non-ideal junctions (less than 100%) transmission/reflection due to their finite size (approaching the wavelength of spin waves). The non-reversible losses due to the scattering are defined by the relaxation time $\tau$, which describes the combined effect of all scattering events. It defines the time required for the amplitude of the propagating spin wave to decrease to 1/e (0.368) of its initial value. The characteristic relaxation time $\tau$ at room temperature is about a nanosecond for conducting ferromagnetic materials (e.g. NiFe), and may be longer for non-conducting materials (e.g. YIG). Another source of dissipation is the scattering from the cross-junctions. In Fig. 2, we presented the results of numerical simulations showing 100% reflection or transmission. These results are valid for the ideal cross-junction comprising two linear chains of spins with just one spin at the intersection point. In any realistic structure, two interfering waves will always have a phase distribution across the junction, which will restrict the total transmission or reflection. The fraction of the reflected/transmitted energy is defined by the ratio of the junction dimension and the wavelength of the interfering waves. The wider the cross-junction, the lower is the junction phase selectivity. One should also consider the possible losses arising from the spin wave packet excitation. There are different physical mechanisms for spin wave excitation such as magnetic field produced by an AC electric current, a spin torque device, or a multiferroic structure. According to theoretical estimates, the utilization of the multiferroics may provide a very high excitation efficiency $\beta=E_{sw}/E_{in} \sim 0.97$ meaning that only 3% of the input power will be dissipated [14]. Taking into account all loss mechanisms, we can estimate the energy dissipated per operation as follows:

$$E_{diss} = N \times E_{sw}\left[\frac{1-\beta}{\beta} + (1-\gamma)^{Nc} \times (1-\exp[-2t/\tau])\right], \tag{4}$$

where $N$ is the number of spin wave packets per circuit, $E_{sw}$ is the energy of the spin wave packet, $\gamma$ is the junction selectivity: $\gamma=(T_0 - T_\pi)/(T_0 + T_\pi)$, $T_0$ and $T_\pi$ are the transmission coefficients of the junction for the 0 and π relative phases of the interfering packets (e.g. $T_0=1$, $T_\pi=0$ for the ideal junction), $Nc$ is the number of cross-junctions per circuit, $t$ is the propagation time, and $\tau$ is the relaxation time for the given material and temperature. The first term in the brackets accounts for the energy losses due to spin wave excitation. The second term in the brackets accounts for the efficiency of the cross-junction and the losses inside the waveguides during the packet propagation. The smaller the junction, the higher is the junction selectivity $\gamma$ at given wavelength. The energy of the spin wave packet $E_{sw}$ should be above the thermal energy $kT$, to ensure reliable output detection. Taking $E_{sw}=20kT$ as a reasonable value, we can estimate the dissipated energy per operation for the circuit shown in Fig.3 (*N=2, β=0.97, T₀=0.9, T_π=0.1, Nc=1, τ=1ns*). The propagation time $t$ can be found as follows: $t=L/v_g$, where $L$ is the length of the input-output path, and $v_g$ is the spin wave group velocity (e.g. $v_g=10^6$ cm/s for magnetostatic surface spin waves in permalloy). For the 100nm long gate, one obtains 10ps for the propagation time. Then, the total energy dissipated per operation is about *8kT*, which is mainly due to the cross-junction non-ideality (*γ=0.8*). On the one hand, there are no fundamental restrictions on improving the junctions and enhancing the conversion efficiency in order to make *β* and *γ* close to 1. On the other hand, there are fundamental limits on the minimum losses due to spin wave propagation. It is not possible to prevent spin wave from scattering on phonons and magnons at any finite temperature. It is interesting to note that the shorter the structure, the less time it takes for the packets to travel, and the less energy will be dissipated inside the circuit. The latter leads us to the important observation: *the energy dissipation decreases for faster switching*. This conclusion differs drastically from the general trend inherent to other logic circuits. For example, the dissipated power increases for faster switching in transistor-based circuits [15] and magnetic cellular automata [16]. The reason for this unusual trend is due to the fact that the proposed circuit operates out of thermal equilibrium [17]. The ratio $t/\tau$ indicates how far the operation regime is from the thermal equilibrium. The lower the ratio $t/\tau$, the less the dissipated energy is. The circuit shown in Fig. 3 would dissipate just *0.8kT* per operation at room temperature if we take into account only the propagation losses ($t/\tau=10ps/1ns=0.01$).

As seen from Eq.(4), the high selectivity of magnetic cross-junction is one of the key requirements for the low-power operation as the overall losses increases exponentially with the number *Nc* of cascaded junctions in the circuit. Recently reported experimental data on spin wave propagation through the magnetic cross-junction structure have shown a pretty high selectivity ($T_0/T_\pi \sim 70$, $\gamma \sim 0.972$) for the micrometer scale structure made of permalloy operating at room temperature and the frequency of 8GHz [18]. Further improvement is expected for nanometer scale junctions by decreasing the width-to-wavelength ratio. Reliable magnetization control via the magneto-electric coupling in composite multiferroics [19] has recently been demonstrated. The electric field required for magnetization rotation (and consequently spin wave excitation) is about 1MV/m [19]. This translates into low energy consumption for spin wave excitation and confirms the theoretical estimates on the high conversion efficiency using multiferroics. All these experimental data support practical feasibility of the proposed concept.

There are concerns about the possibility of cascading the reversible magnonic gates in large scale circuits. The obvious problem with the present design is the lack of unidirectional signal propagation. The example logic gate shown in Fig. 2 has only two ports, where the input signals return to the point of excitation. It would be very difficult to integrate several of such gates in one circuit. As a possible solution, we consider a modified structure shown in Fig. 4. This is a four terminal device with two input ports (A and B) and two output ports (A' and B'). There is a specially designed configuration of phase shifters to prevent signal reflection back to the input ports. There are two possible scenarios for signal propagation: logic input A is the same as the logic input B (e.g. 0,0 or 11). In this case, the packets interfere constructively and propagate towards the output ports without the backscattering. In the second possible case A≠ B (e.g. 01 or 10), there will be multiple reflections from the cross-junctions before the packets accumulate the additional phase shift and reach the output ports. Although the use of the additional phase shifters may resolve the problem and guarantee the unidirectional signal propagation, it significantly complicates the structure.

There are several important questions we leave without an answer: Is it possible to design the Toffoli gate by using only cross-junctions and phase shifters? How to synchronize the outputs of two or several logic gates as the propagation time depends on the logic input? Is it possible to build more than two input/two output logic gates by exploiting the cross-junctions? Each of these questions deserves a more detailed study. The proposed concept may be developed in many different ways by introducing additional components or by designing a more functional logic gates. The described cross-junction can be also considered as a wave-based switch for building conventional logic gates as AND or NOR. The same approach may be applied to other wave-based devices. In this work, we just introduce the concept of the reversible magnonic logic circuits and outline the most intriguing advantages of this approach (e.g. power dissipation decrease for faster switching).

## V. Conclusions

We presented the concept of reversible magnonic logic gates, where logical 0 and 1 are encoded into the phase of the propagating spin wave packets. The input and the output ports are connected by the magnetic waveguides with incorporated cross-junctions and phase shifters. The particular input-output trajectory for each packet depends on the logic input: the relative phase of the propagating spin wave packets. Cross-junction is the key element allowing re-direction of the propagating packets among possible trajectories. The operation of the cross-junction is illustrated by numerical modeling. We present an example of the time-reversible two-input and two-output logic gate comprising one cross-junction and one π-phase shifter. The unique property of the proposed gate is that the power dissipation decreases for faster switching, which is due to the fact that gates operate out of thermal equilibrium and there are no relaxation processes involved in the operation. There are multiple open questions related to the practical feasibility of the proposed approach and the possibility of building complete logic circuitry by using only cross-junctions and phase-shifters. With all pros and cons, we consider the proposed concept as an intriguing alternative to the conventional transistor-based logic, which may potentially lead to a new class of logically reversible logic circuits with ultra-low power consumption.

## Acknowledgments

We would like to thank Dr. Dmitri E. Nikonov (Intel Co) for valuable discussions. The work was supported by the Nanoelectronics Research Initiative (NRI) (Dr. Jeffrey J. Welser, NRI Director) via the Western Institute of Nanoelectronics (WIN, director Dr. Kang L Wang).

**Figure Captions**

Fig.1 (A) Schematics of a reversible magnonic logic gate. Logical 0 and 1 are encoded by the relative phases of propagating spin wave packets with respect to a reference packet. The packets propagate through the structure consisting of waveguides, cross-junctions, and π-phase shifters. There are several possible trajectories for each packet to travel from the input to the output. Some of the trajectories contain phase shifters and some of them do not. The particular trajectory depends on the combination of the input phases. (B) Examples of the Buffer and Not gates, which contain two possible trajectories for the propagating packets without and with the phase inverter, respectively.

Fig.2 (A) Schematics of the magnonic cross-junction represented by two crossing chains of spins. The spins in the chains are coupled via the exchange interaction. The input spin wave packets are excited by perturbing the edge spins along the X axis. The relative phase (0 or π) of the excited spin wave depends on the direction of the edge spin's perturbation. (B) Results of numerical simulations of packet propagation through the cross-junctions. Two packets propagate through the junction without any reflection if the relative phase difference is 0. The packets are reflected from the junction if the relative phase is π.

Fig.3 An example of two-input two-output reversible logic gate. The gate consists of one junction and one phase shifter. Two ports A and B serve as both input and output terminals. The spin wave packets excited at ports A and B have two possible trajectories of propagation depending on the relative phase. In case A≠B, the packets are reflected from the junction and return to the inputs with the same initial phase. In case A=B, the packets will propagate through the junction. The packet traveling to port A will obtain an additional π-phase shift. The truth table illustrates the logic operation (logical 0 and 1 corresponds to the initial phases 0 and π, respectively).

Fig.4. A modified version of the logic gate with unidirectional signal propagation. In case of the backward reflection from the junction, the additional phase shifters provide a phase shift to the reference signal. As a result of the multiple scatterings, the wave propagates through the gate towards the output (from the left to the right).

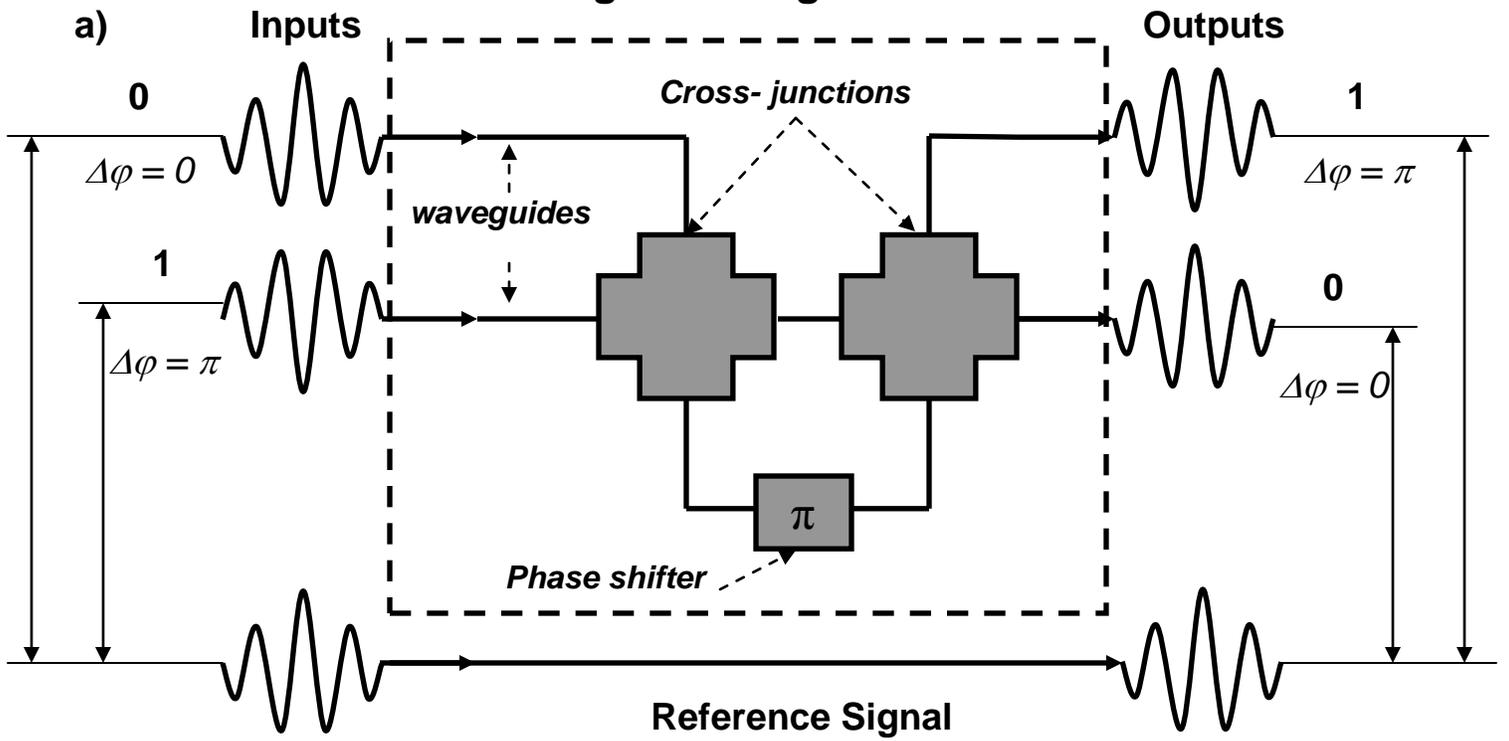

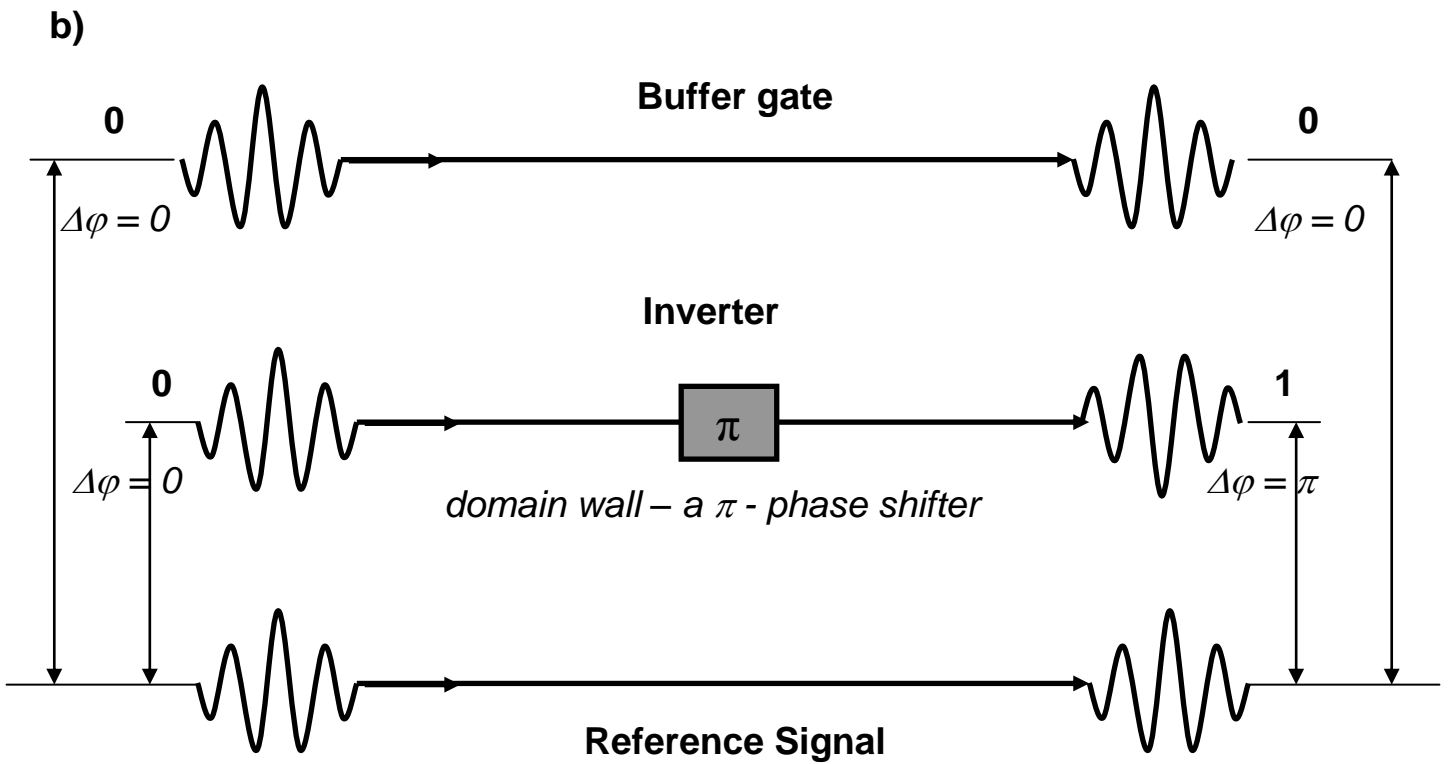

**Fig.1**

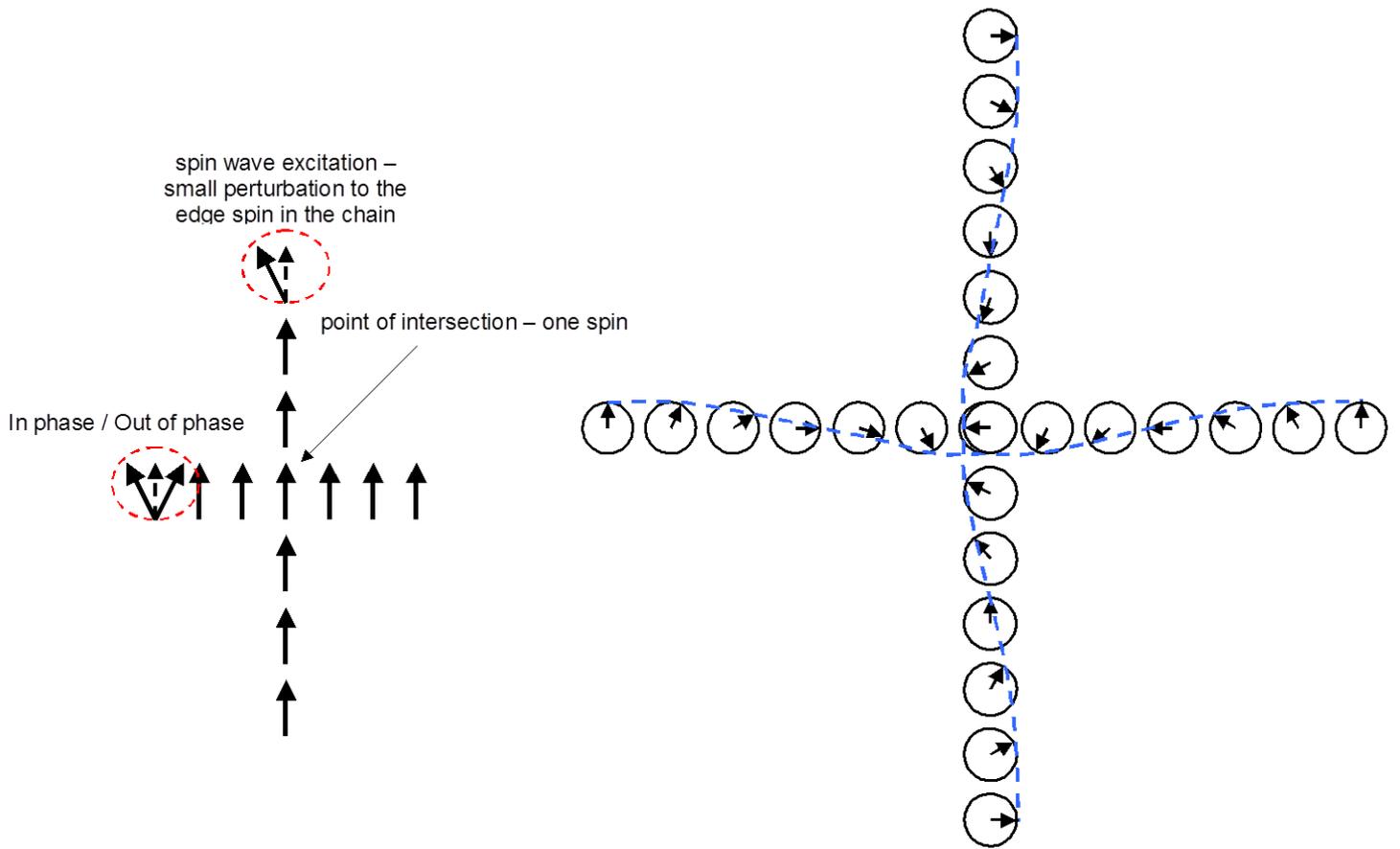

## Two waves are in phase: 100% Transmission

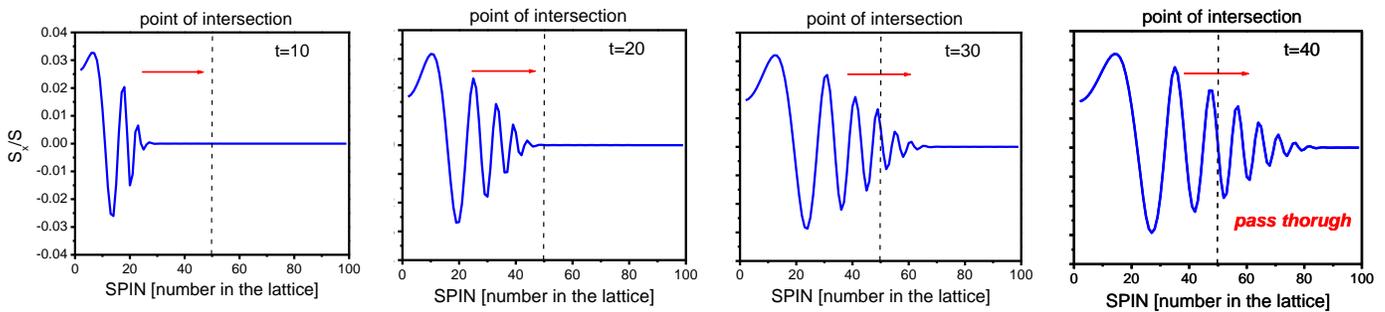

## Two waves are out of phase: 100% Refraction

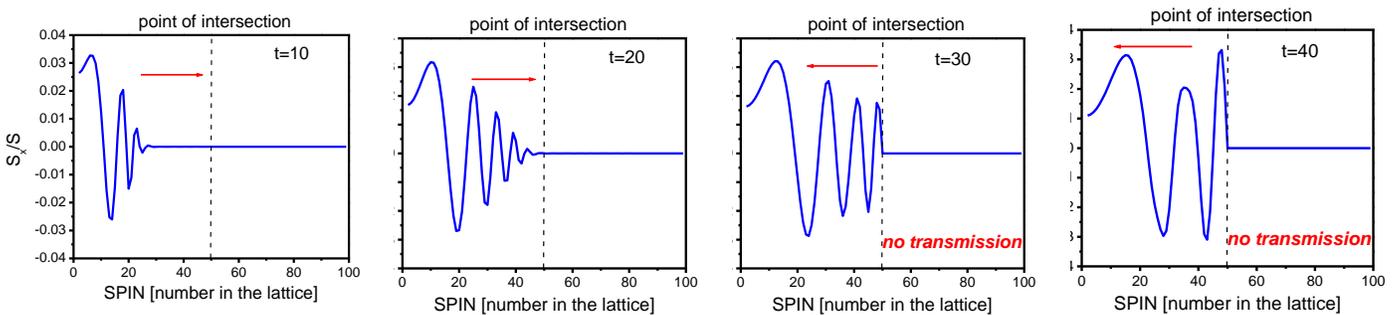

**Fig.2**

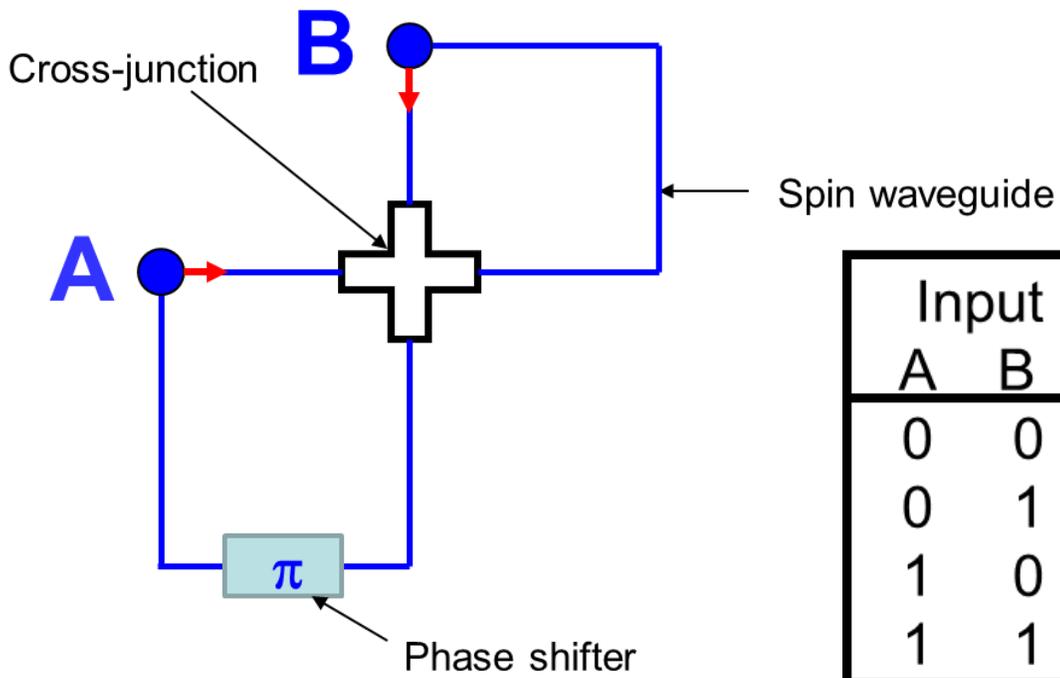

| Input | | Output | |
|---|---|---|---|
| A | B | A | B |
| 0 | 0 | 1 | 0 |
| 0 | 1 | 0 | 1 |
| 1 | 0 | 1 | 0 |
| 1 | 1 | 0 | 1 |

A ≠ B  100% Reflection

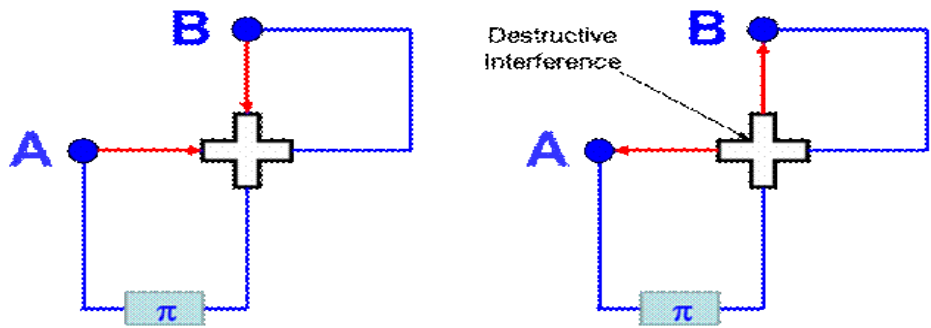

A = B  100% Transmission

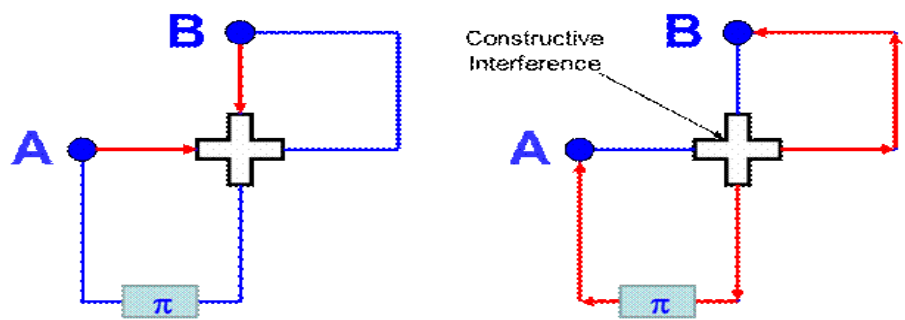

**Fig.3**

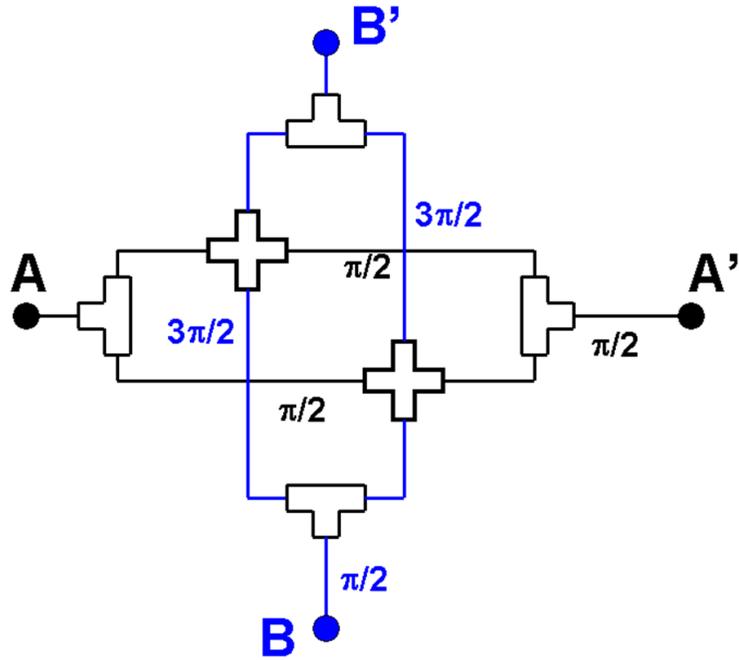

Case #1: A=B waves come through each junction without reflection

Case #2: A≠B there are multiple reflections before waves will come through

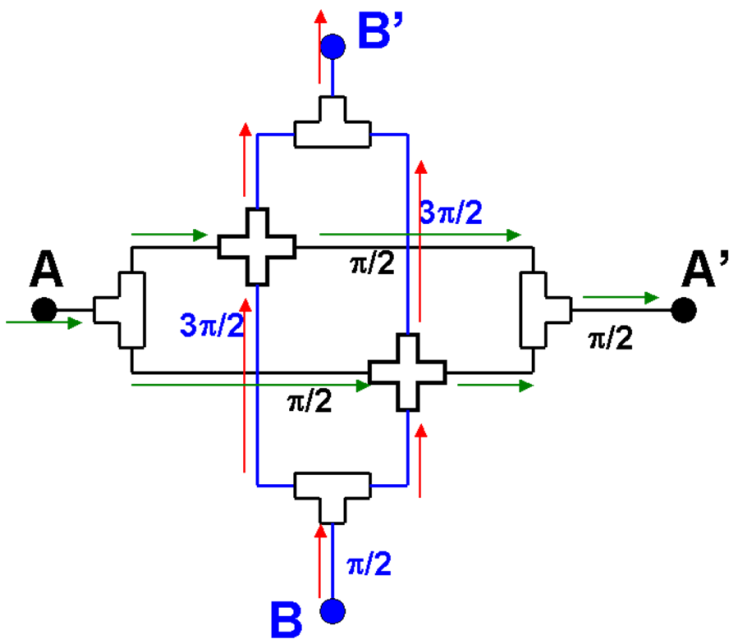
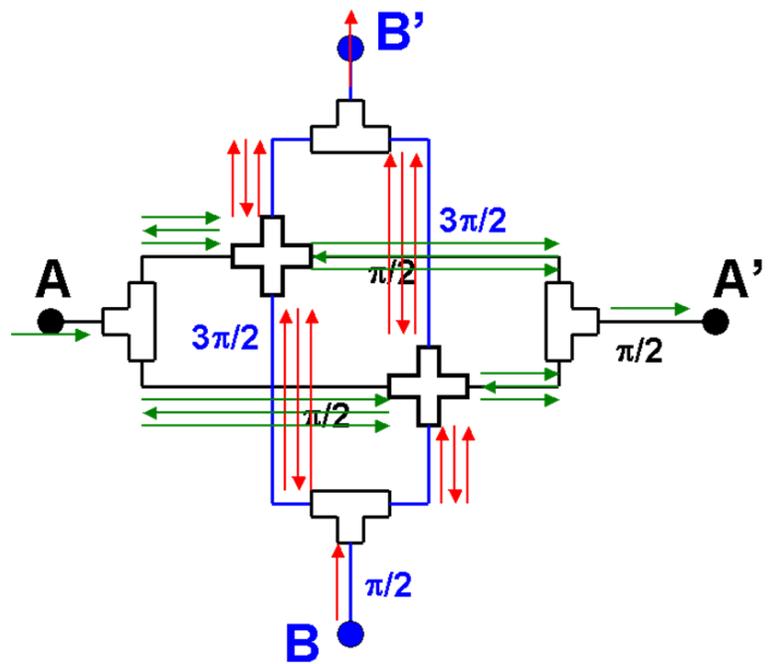

**Fig.4**